\documentclass[conference,a4paper]{APSIPA2019}
\usepackage{multirow}
\usepackage{graphicx}
\usepackage{float}

\usepackage{amsmath}
\usepackage[psamsfonts]{amssymb}
\usepackage{amsxtra}
\usepackage{threeparttable}
\usepackage{booktabs}
\usepackage[colorlinks=true,linkcolor=black,anchorcolor=black,citecolor=black,filecolor=black,menucolor=black,runcolor=black,urlcolor=black]{hyperref}

\begin{document}
\title{Emotion Dependent Facial Animation from Affective Speech}

\author{%
\authorblockN{%
 Rizwan Sadiq, Sasan Asadiabadi and Engin Erzin
}
\authorblockA{%
Multimedia, Vision and Graphics Laboratory,\\
College of Engineering, Ko\c{c}¸ University, Istanbul, Turkey \\
E-mail: rsadiq13, sabadi15, eerzin@ku.edu.tr}
}

\maketitle
\thispagestyle{empty}

\begin{abstract}
  In human-to-computer interaction, facial animation in synchrony with affective speech can deliver more naturalistic conversational agents. In this paper, we present a two-stage deep learning approach for affective speech driven facial shape animation. In the first stage, we classify affective speech into seven emotion categories. In the second stage, we train separate deep estimators within each emotion category to synthesize facial shape from the affective  speech. Objective and subjective evaluations are performed over the SAVEE dataset. The proposed emotion dependent facial shape model performs better in terms of the Mean Squared Error (MSE) loss and in generating the landmark animations, as compared to training a universal model regardless of the emotion.
\end{abstract}

\section{Introduction}
In the past few years, deep learning algorithms have successfully deployed in many areas of research, such as computer vision and pattern recognition, speech processing, text-to-speech and many other machine learning related areas \cite{krizhevsky2012imagenet,bahdanau2014neural,amodei2016deep}. This has led the gradual performance improvement of many deep learning based statistical systems applied to various problems. These fundamental improvements also motivated researchers to explore problems of human-computer interaction (HCI), which have long been studied. One such problem involves understanding human emotions and reflecting them through machines, such as emotional dialogue models \cite{zhou2018emotional,huang2018automatic}.

Its natural for humans to identify the emotions and react accordingly. However, perception of emotion and affective response generation are still challenging problems for machines. In this study, we  target to estimate facial animation representations from affective speech. For this purpose, we construct a two stage process, where the first stage uncovers the emotion in the affective speech signal, and the second stage defines a mapping from the affective speech signal to facial movements conditioned on the emotion.  



Emotion recognition from speech has been an active research area in HCI for various applications, specially for humanoid robots, avatars, chat bots, virtual agents etc. Recently, various deep learning approaches have been utilized to improve emotion recognition performance, which includes significant challenges in realistic HCI settings.
Before the deep learning era, classical machine learning models, such as hidden Markov
models (HMMs), support vector machines (SVMs), and decision tree-based methods, have been used in speech emotion recognition \cite{pan2012speech,schuller2003hidden,lee2011emotion,sadiq2017affect}.
As deep neural networks become widely available and computationally feasible, wide range of studies adapted complex deep neural network models for the speech emotion recognition problem. Among these models, 
Convolutional neural network (CNN) based models are successfully utilized for improved speech emotion recognition  \cite{bertero2017first,badshah2017speech}. In another setup, a recurrent neural network model based on text and speech inputs has been successfully used for emotion recognition \cite{yoon2018multimodal}. 
Early research on acoustic to visual mappings investigated different methods including Hidden Markov Models (HMM), Gaussian Mixture Models (GMM) and Dynamic Bayesian Networks (DBN).  
One of the pioneers to produce speech driven facial animation used classical HMM \cite{yamamoto1997speech}, where they successfully mapped states of the HMM to the lip parameters; moreover they also proposed to utilize visemes in mapping HMM states to visual parameters. Their idea of using visemes was later used by many researchers for synthesizing facial animations via speech signals  \cite{bozkurt2007comparison,verma2003using}. 

Another early work on facial animation used classical machine learning approaches, where the 3-D facial movements were predicted from the LPC and RASTA-PLP acoustic features \cite{brand1999voice}. Later, \cite{kakumanu2001speech} proposed to consider context by tagging video frames to audio frames from past and future. 
Mostly, the mapping from speech to visual representations is performed over offline data. As a real-time solution, \cite{vougioukas2018end} proposed a generative adversarial network (GAN) based audio to visual mapping system that can synthesize the visual sequence at 150~frames/sec. 

Recently,  \cite{taylor2017deep} focused on generating facial animations solely from the phoneme sequence.  They use a DNN based sliding window predictor that learns arbitrary nonlinear mappings from phoneme label input sequences to facial lip movements. Contrary to the conventional methods of mapping phoneme sequence to fixed number of visemes in \cite{verma2003using,bozkurt2007comparison}, they generate a sequence of output video frames for the input speech frame, and took the mean of that sequence to map the active speech frame to a single video frame.

In the literature, research on facial animation from affective speech is limited, this is mostly due to scarcity of labeled affective audio-visual data. 
In a recent study, the problem of embedding emotions in speech driven facial animations has been modeled through an LSTM model \cite{pham2017speech}. They employed the RAVDESS dataset in their study \cite{livingstone2012ravdess}, which only includes two-sentence setup with limited phonetic variability.
In our earlier work, \cite{asadiabadi2018multimodal}, we proposed a deep multi-modal framework, combining the work of \cite{taylor2017deep} using phoneme sequence with spectral speech features to generate facial animations. Our work successfully demonstrated the good use of CNN models to capture emotional variability in mapping affective speech to facial representations.

The IEMOCAP dataset \cite{busso2008iemocap} has been widely used in the literature for audio-visual emotion recognition task. Although the IEMOCAP delivers a rich set of affective audio-visual data, it mostly lacks frontal face videos and emotion categories are not all balanced. In this study, we choose to use the SAVEE \cite{cooke2006audio} dataset. It delivers a balanced set of affective data as well as including clear frontal face videos, which help significantly to train better facial representation models.

Our contributions in this paper are two fold. First we present a speech emotion recognition system which is trained with the SAVEE dataset to understand the emotional content in underlying speech signal. Secondly, we present a  emotion dependent speech driven facial animation system which map the speech signal to facial domain according to emotional content.

The reminder of this paper is organized as follows. In Section~\ref{sec:Method}, we describe the proposed methodology for the emotions based speech driven facial shape animations. we give experimental evaluations in Section~\ref{sec:Res}. Finally, conclusions are discussed in Section~\ref{sec:conc}.

\begin{figure}[t]
\centering
    \includegraphics[width=90mm]{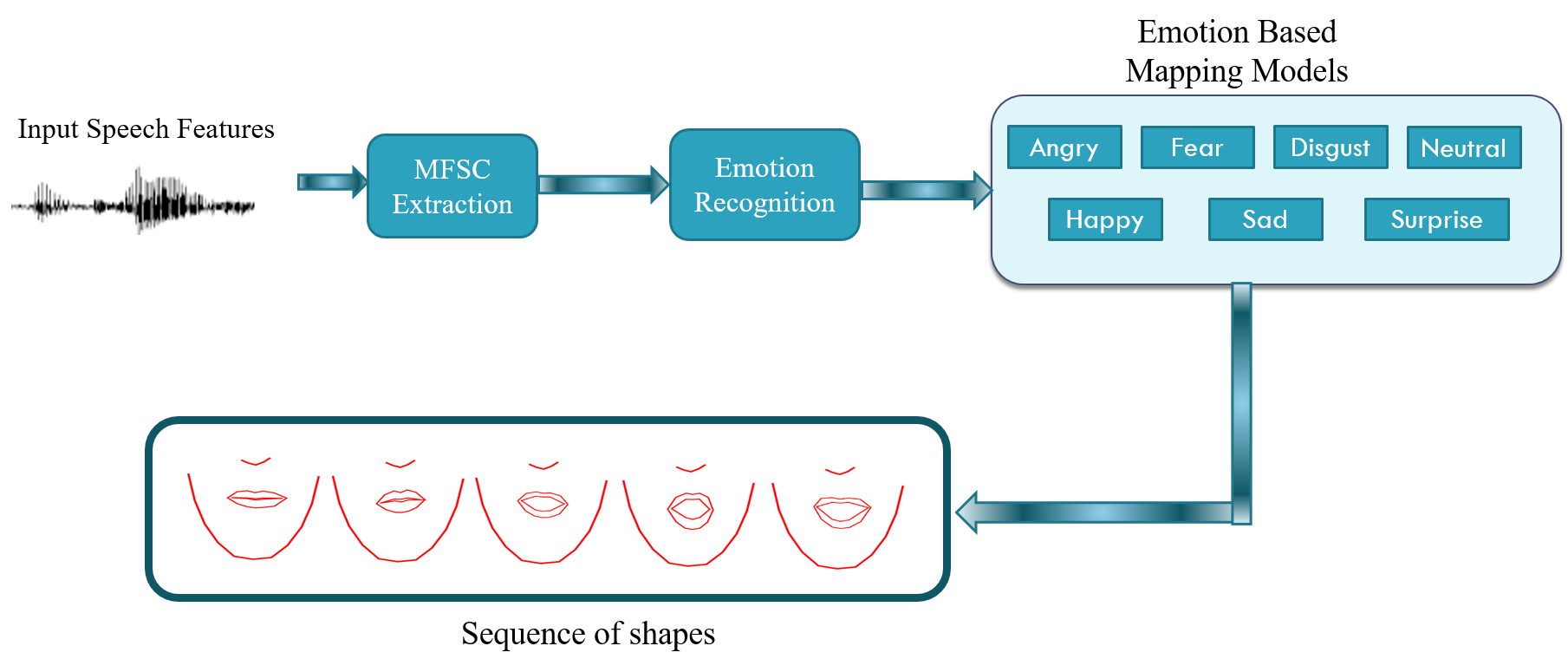}
    \caption{Block diagram of the proposed emotion dependent facial animation system.}
    \label{fig:1}
\end{figure}
\section{Methodology}
\label{sec:Method}
We model the affective speech animation problem as a cascade of emotion classification followed by facial parameter regression as depicted in Figure~\ref{fig:1}.  The affective content of the input speech is first classified into one of the 7 emotion categories, then the corresponding emotion dependent deep model maps the spectral speech representation to the visual shape parameters. 

\subsection{Dataset}
In this study, we use the Surrey Audio-Visual Expressed Emotion (SAVEE) dataset to evaluate the affective speech animation models \cite{cooke2006audio}. The  dataset consists of video clips from 4 British male actors with six basic emotions (disgust, anger, happy, sad, fear surprise) and the neutral state. A total of 480 phonetically balanced sentences are selected from the standard TIMIT corpus \cite{garofolo1993darpa} for every emotional state. Audio is sampled at 44.1~kHz with video being recorded at a rate of 60~fps. Phonetic transcriptions are provided with the dataset. A total of approximately 102~K frames are available to train and validate models. Face recordings are all frontal and faces are painted with blue markers for tracking of facial movements.

\subsection{Feature extraction}

\subsubsection{Acoustic features}
\begin{figure}[tb]
\centering
    \includegraphics[width=90mm]{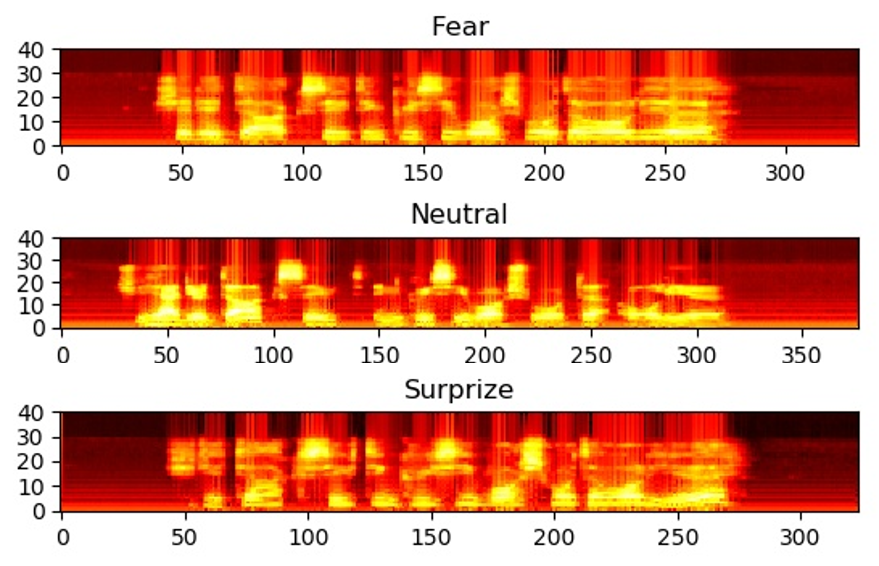}
    \caption{MFSC spectrograms of one sample sentence, "she had your dark suit in greasy wash water all year," with the fear, neutral and surprise affective states.}
    \label{fig:spectros}
\end{figure}

We use the mel-frequency spectral coefficients, aka, MFSC to represent the speech acoustic features. For each speech frame, 40 dimensional MFSC features are extracted to define the acoustic energy distribution over 40 mel-frequency bands. Python's speech feature library is used for the feature extraction. The MFSC features are extracted from pre-emphasized overlapping Hamming windowed frames of speech at 100~Hz. The extracted feature set is z-score normalized to have zero mean and unit variance in each feature dimension. We represent the set of acoustic feature vectors as $\left\{f^a_j\right\}^N_{j=1}$, where $f^a_{j} \in \mathbb{R}^{40 \times 1}$ and $N$ is the total number of frames. Figure~\ref{fig:spectros} presents sample MFSC spectrograms from three different emotions.

A temporal sliding window of spectral image is defined to capture the spatio-temporal characterization of the input speech as $F^a_j = [f^a_{j-\Delta_{a}},...,f^a_j,...,f^a_{j+\Delta_{a}}]$, where $F^a_j$ is a $40 \times K_{a}$ image, $K_{a}=2\Delta_{a}+1$ is the temporal length of the spectral image at time frame $j$ and the sliding window moves with stride~$1$ for $j=1, \dots, N$. 


\subsubsection{Facial shape features}
\begin{figure}[t]
\centering
    \includegraphics[width=60mm]{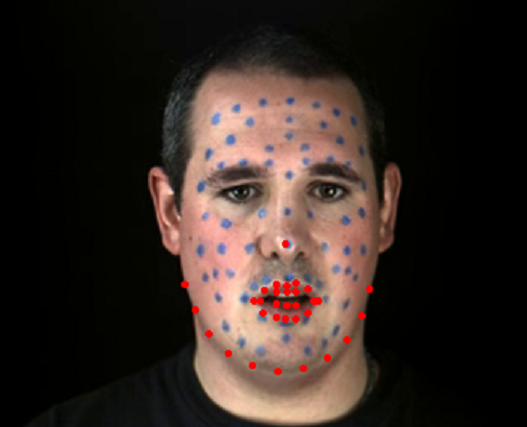}
    \caption{Facial markers on the face: Blue markers are from the SAVEE dataset, red markers are the extracted landmarks using the Dlib face detector.}
    \label{fig:marks}
\end{figure}

As defined in our previous work \cite{asadiabadi2018multimodal}, the facial shapes are described with a set of $M=36$ landmark points on the lower face region, along the jaw line, nose, inner and outer lips and represented as $S_{j} = \left\{(x^{j}_{i},y^{j}_{i})\right\}^M_{i=1}$, where $i$ is the landmark index and $j$ is the sample index. The landmarks used in our experiments are extracted using the Dlib face detector \cite{king2009dlib} shown as red dots in Figure~\ref{fig:marks}. To obtain a one-to-one correspondence between the acoustic and visual features, the videos in the dataset are re-sampled to 25~fps. From the re-sampled videos, the landmark points are extracted at a 25~Hz rate and later up-sampled to 100~Hz using cubic interpolation, thus yielding a one-to-one match to the acoustic feature sequence.

The extracted facial shape set 
$S=\left\{S_j\right\}^N_{j=1}$
are aligned using the Generalized Procrustes Analysis (GPA) \cite{ref:procrutes} to remove the possible rotation, scale and position differences across speakers. Then a statistical face shape model is generated utilizing Principal Component Analysis (PCA) algorithm. PCA projects the shapes in the facial space $S$ to a lower dimensional uncorrelated parameter space 
$P=\left\{P_j\right\}^N_{j=1}$.
The projection between the two spaces is carried out using the mean and truncated eigenvector matrix of the covariance of $S$ as defined in detail in \cite{asadiabadi2018multimodal}. In this study, $18$ PCA parameters are used, which are covering around $99 \%$ of the variation in the dataset. 

The target output sequence of the DNN is obtained utilizing a sliding window of size $K_{v}$ and stride $1$ over the shape parameter space. The temporal shape feature sequence is represented as $\left\{F^v_j\right\}^N_{j=1}$ where  $F^v_j = [f^v_{j-\Delta_{v}},...,f^v_j,...,f^v_{j+\Delta_{v}}] \in \mathbb{R}^{18K_{v} \times 1}$ with $K_{v}=2\Delta_v + 1$.

\subsection{Speech emotion recognition}
Emotion recognition from speech has been widely studied in the literature and the speech spectrogram is known to discriminate emotions well. As Figure~\ref{fig:spectros} demonstrates sample variations of the MFSC spectrograms on three different emotions, we set the $F^a_{j}$ spectral image as the acoustic feature to train and predict the emotion of the underlying speech. To this effect, we use a deep emotion recognition network (DERN), which includes 3 convolutional layers followed by a fully connected layer and a softmax layer at the output. The details of the DERN network are given in Table~\ref{tab:emotion}. 

In the emotion recognition phase, the DERN outputs an emotion label for each frame in a given test utterance as, $\tilde{e}_j=DERN(F^a_{j})$, where $\tilde{e}_j$ is the estimated emotion label for frame $j$. For an utterance consisting of $T$ frames, the estimated emotion sequence can be given as $\{\tilde{e}_j\}_{j=1}^T$. Let's index the 7 emotions used in this study as $e^1, e^2, \dots, e^7$. Utterance level emotion probability for the $i$-th emotion can be defined as
\begin{equation}
    p_i = \frac{1}{T} \sum_{j=1}^T 1(e^i = \tilde{e}_j),
\end{equation}
where ones function, $1()$, returns 1 when condition is true else a zero. Then the top two emotions for the utterance can be identified as
\begin{equation}
    i^* = \arg\max_i p_i \;\; \text{and}\;\; i^{**} = \arg\max_{i - \{i^*\}} p_i.
\end{equation}

The top second emotion is utilized when the confidence to the top emotion is weak. Hence, probability of the top second emotion is updated as
\begin{equation}
  p_{i^{**}} =
  \begin{cases}
    0 & \text{if $p_{i^{*}} > 0.65$} \\
    p_{i^{**}} & \text{otherwise}.
  \end{cases}
\end{equation}
Then the normalized probabilities for the utterance level top two emotions are set as
\begin{equation}
    p^* = \frac{p_{i^{*}}}{p_{i^{*}}+p_{i^{**}}} \;\; \text{and}\;\; p^{**} = \frac{p_{i^{**}}}{p_{i^{*}}+p_{i^{**}}},
\end{equation}
corresponding to the top two emotions $e^*$ and $e^{**}$, respectively.
\begin{table}[bht]
\caption{Network architecture for the DERN}
\label{tab:emotion}
\centering
\scalebox{0.95}{
\begin{tabular}{l c c c c c c}
\toprule[1pt]\midrule[0.3pt]
\textbf{Layer} & \textbf{Type} & \textbf{Depth} & \textbf{Filter Size} & \textbf{Stride}\\
\midrule
\hline
1 & CONV+ReLu & 32 & 5x5 & - \\
\hline
2 & MaxPooling & 32 & 3x3 & 2 \\
\hline
3 & CONV+ReLu & 64 & 5x5 & - \\
\hline
4 & MaxPooling & 64 & 3x3 & 2 \\
\hline
5 & CONV+ReLu & 128 & 5x5 & - \\
\hline
6 & MaxPooling & 128 & 3x3 & 2 \\
\hline
7 & FC+ReLu & 256 & - & - \\
\hline
8 & Dropuout & - & - & - \\
\hline
9 & FC+Softmax & 7 & - & - \\
\hline
\end{tabular}}
\end{table}

\subsection{Emotion dependent facial shape regression}
\label{ref: av2}
We train a deep shape regression network (DSRN) to estimate the facial shape features $F^v_j$ from the acoustic MFSC spectrogram images $F^a_j$ for each emotion category, separately. Note that each DSRN model is trained for each emotion category, hence the estimated facial shape feature can be defined as the fusion of two estimates extracted with the top two emotions,
\begin{equation}
    \label{equ:two label}
    \tilde{F}^v_j = p^* DSRN(F^a_j | {e}^*) + p^{**} DSRN(F^a_j | {e}^{**}).
\end{equation}

In this formalism, we have $K_v$ number of facial shape estimates for the frame $j$ that are extracted from the neighboring estimates. The final facial shape estimation is defined as the average of $K_v$ estimates,
\begin{equation}
  \hat{f}^v_j =  \frac{1}{K_v} \sum_{i=j-\Delta_v}^{j+\Delta_v} \tilde{f}^v_i .  
\end{equation}

The DSRN is constructed with 4 convolutional layers, which are followed by 2 fully connected layers to estimate the shape features. We use dropout regularization method with probability 50\% in the fully connected layers only, to overcome the over-fitting. ReLu activation function is used at each layer with Adam optimizer for hyper learning rate optimizations. Mean square error (MSE) is chosen as the objective function to be minimized. During the training, we only apply convolutions and pooling over the frequency axis, in order to further prevent any over-fitting, and also to preserve the temporal nature of speech. Detailed specification of the DSRN network is given in Table~\ref{tab:2}.
\begin{table}[bht]
\caption{Network architecture for the DSRN}
\label{tab:2}
\centering
\scalebox{0.95}{
\begin{tabular}{l c c c c c c}
\toprule[1pt]\midrule[0.3pt]
\textbf{Layer} & \textbf{Type} & \textbf{Depth} & \textbf{Filter Size} & \textbf{Stride}\\
\midrule
\hline
1 & CONV+ReLu & 32 & 5x1 & - \\
\hline
2 & MaxPooling & 32 & 3x1 & 2x1 \\
\hline
3 & CONV+ReLu & 64 & 5x1 & - \\
\hline
4 & MaxPooling & 64 & 3x1 & 2x1 \\
\hline
5 & CONV+ReLu & 128 & 5x1 & - \\
\hline
6 & MaxPooling & 128 & 2x1 & 2x1 \\
\hline
7 & CONV+ReLu & 128 & 3x1 & - \\
\hline
8 & MaxPooling & 128 & 2x1 & 2x1 \\
\hline
9 & FC+ReLu & 1024 & - & - \\
\hline
10 & Dropuout & - & - & - \\
\hline
11 & FC+ReLu & 500 & - & - \\
\hline
12 & Dropuout & - & - & - \\
\hline
13 & FC+Multi-Reg & $18 \times K_v$ & - & - \\

\hline
\end{tabular}}
\end{table}

%
\section{Experimental results}
\label{sec:Res}

We deploy 5-fold cross-validation training and testing scheme, where 10\% of the dataset is hold for testing the trained networks. From the remain of the dataset 80\% is used for training and the remaining 20\% is used for validation in each fold. Both of the deep models, DERN and DSRN, are trained using the Keras\footnote{\url{https://keras.io/}} with Tensorflow \cite{abadi2016tensorflow} backend on a NVIDIA TITAN XP GPU. The temporal window size for acoustic and visual shape features is set as $K_a = 15$ and $K_v = 5$, respectively. 

\subsection{Objective evaluations}

\subsubsection{Emotion recognition results}

The DERN model is trained over 200 epochs using the categorical cross-entropy loss function. Utterance level emotion recognition performances over the validation set are reported for each emotion category in Table~\ref{tab:3}. All models sustain high average recall rates. The surprise emotion category model is observed to suffer the most, compared to other emotions.

\begin{table}[bht]
\caption{Utterance level accuracy (\%) for the speech emotion recognition in each emotion category}
\label{tab:3}
\centering
\scalebox{0.85}{
\begin{tabular}{ccccccc}
\toprule[1pt]\midrule[0.3pt]
Angry  & Disgust & Fear & Happy & Neutral & Sad & Surprise  \\
\midrule
\hline
 72.15 & 75.38 & 73.21 & 71.71 & 91.93 & 75.19 & 67.98 \\ \hline

\end{tabular}}
\end{table}


\subsubsection{Facial shape regression results}
\begin{figure}[bht]
\centering
    \includegraphics[width=90mm]{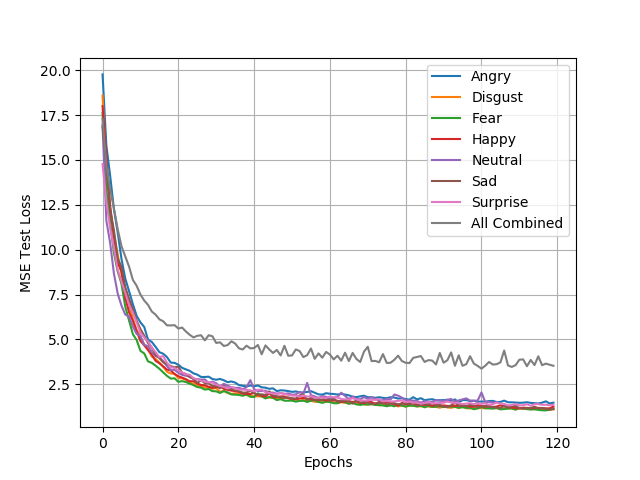}
    \caption{The MSE loss over the validation data along the epochs for the emotion dependent (separate for each emotion) and independent (all combined) models.}
    \label{fig:loss}
\end{figure}

The mean squared error between predicted and original PCA coefficient values is used as the loss function for the training.
The DSRN model is trained for each emotion category separately, which defines the emotion dependent models. We as well train an emotion independent model using all combined training data, which sets a baseline for evaluations. Figure~\ref{fig:loss} presents the MSE loss curve over the validation data through the learning process for emotion dependent and independent models. Note that the proposed emotion dependent regression attains significantly lower MSE loss values, which are more than 65\% reduction in MSE compared to the emotion independent combined model. 

The MSE loss performance of the cascaded trained DERN and DSRN models over the test set is given in Table \ref{tab:4}. As obvious from the table, the proposed cascaded DERN and DSRN scheme performs better than the baseline model in terms of MSE in shape domain. In should be noted that given the true emotion labels of the utterances, the test loss of the DSRN is remarkably lower than the all combined model, which makes room to improve the performance of the DERN module in future work.

\begin{table}[b]
\caption{MSE loss on the test set for cascaded DERN and DSRN, true emotions and DSRN vs all combined model}
\label{tab:4}
\centering
\scalebox{0.85}{
\begin{tabular}{l c c c}
\toprule[1pt]\midrule[0.3pt]
\textbf{Method}  & \textbf{DERN+DSRN}& \textbf{Actual Emo+DSRN} & \textbf{ALL Combined}\\
\midrule
\hline
MSE & {5.57} & {3.23} & {6.73} \\
\hline
\end{tabular}}
\end{table}

\subsection{Subjective evaluations}
The resulting facial shape animations are also evaluated subjectively through visual user preference study. We use a mean opinion score (MOS) test to subjectively evaluate animations of the emotion dependent and independent models. The test is run with 15 participants using 7 conditions, which are the animations of utterances from the 7 emotion categories. Each test session for a participant contains 14 clips, where each condition is tested with the emotion dependent and independent models. During the test, all clips are shown in random order. In the test, each participant is asked to evaluate clips based on synchronization and emotional content of the animations using a five-point preference scale (1: Bad, 2: Poor, 3: Fair, 4: Good, and 5: Very Good).

\begin{table}[t]
\caption{Average preference scores for the emotion dependent and independent (all combined) model animation evaluations}
\label{tab:5}
\centering
\scalebox{0.95}{
\begin{tabular}{l c c}
\toprule[1pt]\midrule[0.3pt]
\textbf{Method} & \textbf{Mean} & \textbf{Std}  \\
\midrule
\hline
Emotion Dependent & {3.03} & 0.96 \\
\hline
All Combined & 2.74 & 0.88\\
\hline
\end{tabular}}
\end{table}

\begin{figure}[bht]
\centering
    \includegraphics[width=90mm]{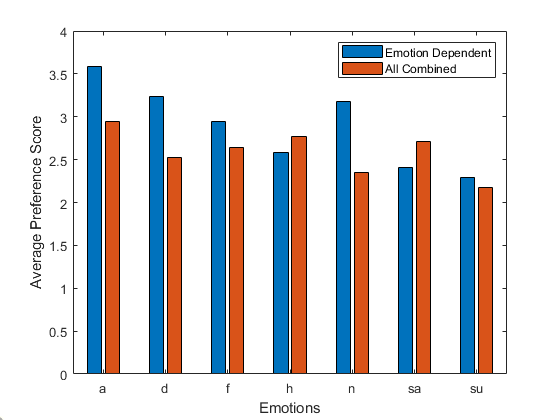}
    \caption{Emotion category based average preference scores for the emotion dependent and independent (all combined) model animation evaluations. }
    \label{fig:subjec}
\end{figure}

The average preference scores are listed in Table~\ref{tab:5}. The proposed emotion dependent facial shape animation scheme is preferred over the baseline emotion independent scheme. Furthermore, in each emotional category a similar preference tendency, except for happy and sad categories, is observed as presented in Figure~\ref{fig:subjec}. 


%
\section{Conclusions}
\label{sec:conc}
In this study we propose an emotion dependent speech driven facial animation system. A statistical shape model (PCA based) is trained to project the shape data into an uncorrelated lower dimensional parameter space, capturing 99\% of variation in the training data. We observe that training separate models to map acoustic spectral features to visual shape parameters performs better than training a universal network with all the emotions combined. Our proposed emotion dependent facial shape animation model outperforms the emotion independent universal model in terms of the MSE loss and also in the subjective evaluations, facial animations of the proposed method preferred higher.
As a future work we will investigate the ways to improve the accuracy of the speech emotion recognition system.

\section{Acknowledgements}
This work was supported in part by the Scientific and Technological Research Council of Turkey (T\"{U}B\.{I}TAK) under grant number 217E107.

\bibliographystyle{unsrt}
\bibliography{main.bib}
\end{document}